\documentclass[journal]{IEEEtran}
\usepackage{amsmath,amssymb}
\usepackage{algorithm}
\usepackage{algpseudocode}
\usepackage{cite}
\usepackage{graphicx}
\usepackage{placeins}
\usepackage{tikz} 
\usetikzlibrary{calc}
\usepackage{hyperref}
\usepackage{cleveref}
\usepackage{booktabs}
\usepackage{microtype}
\usepackage{setspace}
\usepackage{xcolor}
\usepackage{algpseudocode} 
\algrenewcommand\algorithmicrepeat{\textbf{Repeat}}          
\algrenewcommand\algorithmicuntil{\textbf{Until}}           
\algrenewcommand\algorithmicreturn{\textbf{Return}}

\begin{document}

	\title{Rotatable Antenna Enabled Covert Communication}

	\author{
	\IEEEauthorblockN{Qi Dai\textit{, Graduate Student Member, IEEE}, Beixiong Zheng\textit{, Senior Member, IEEE}, \\Yanhua Tan\textit{, Graduate Student Member, IEEE}, Weidong Mei\textit{, Member, IEEE}, Shiqi Gong,\\ Jie Tang\textit{, Senior Member, IEEE}, and Chengwen Xing\textit{, Member, IEEE}}
	\thanks{This work was supported in part by the National Natural Science Foundation of China under Grant 62571193 and Grant 62331022; in part by the Guangdong Program under Grant 2023QN10X446 and Grant 2023ZT10X148; and in part by the GJYC Program of Guangzhou under Grant 2024D01J0079 and Grant 2024D03J0006. \textit{(Corresponding author: Beixiong Zheng.)}}
	\thanks{Qi Dai, Beixiong Zheng and Yanhua Tan are with the School of Microelectronics, South China University of Technology, Guangzhou 511442, China (E-mails: 7qidai@gmail.com; bxzheng@scut.edu.cn; tanyanhua06@163.com).}
	\thanks{Weidong Mei is with the National Key Laboratory of Wireless Communications, University of Electronic Science and Technology of China, Chengdu, 611731, China (E-mail: wmei@uestc.edu.cn).}
    \thanks{Shiqi Gong and Chengwen Xing are with the School of Information and Electronics, Beijing Institute of Technology, Beijing 100081, China (E-mails: gsqyx@163.com; xingchengwen@gmail.com).}
    \thanks{Jie Tang is with the School of Electronic and Information Engineering, South China University of Technology, Guangzhou 510640, China (E-mail: eejtang@scut.edu.cn).}
}
	\maketitle
	\begin{abstract}
		Unlike conventional fixed-antenna architectures, rotatable antenna (RA) has shown great potential in enhancing wireless communication performance by exploiting additional spatial degrees of freedom (DoFs) in a cost-effective manner. In this letter, we propose a novel RA-enabled covert communication system, where an RA array-based transmitter (Alice) sends covert information to a legitimate user (Bob) in the presence of multiple wardens (Willies). To maximize the covert rate, we optimize the transmit beamforming vector and the rotational angles of individual RAs, subject to the constraints on covertness, transmit power, and antenna rotational range. To address the non-convex formulated problem, we decompose it into two subproblems and propose an efficient alternating optimization (AO) algorithm to solve the two subproblems iteratively, where the second-order cone programming (SOCP) method and successive convex approximation (SCA) approach are applied separately. Simulation results demonstrate that the proposed RA-enabled covert communication system can provide significantly superior covertness performance to other benchmark schemes.
	\end{abstract}


	\section{Introduction}
	Over the past decades, covert communication, also known as low-probability-of-detection (LPD) communication, has emerged as a cutting-edge security paradigm in modern wireless networks, driven by the increasing demand for secure and undetectable data transmission in the presence of adversaries. Different from conventional secure technologies such as physical layer security (PLS) and upper layer encryption, which focus on protecting the content of the messages, covert communication seeks to conceal the very existence of the communication itself, making it an essential technique for corporate finance, military operations, and privacy-sensitive civilian communication\cite{Bash}. To improve the feasibility and performance of covert communications, several techniques have been explored, including uncertainty exploitation\cite{Noise,channel}, and artificial noise jamming\cite{AN}. Nevertheless, existing methods are primarily based on fixed-antenna systems, in which the spatial positions and orientation of antennas remain unchanged. This fundamental limitation severely hinders the exploration of spatial degrees of freedom (DoFs), thereby constraining system spectrum efficiency and covertness performance. Thus, the development of more flexible antenna architectures is highly imperative for enabling resilient and stealthy wireless transmissions.

	\begin{figure}[!t] 
	\centering
	\includegraphics[width=2.5in]{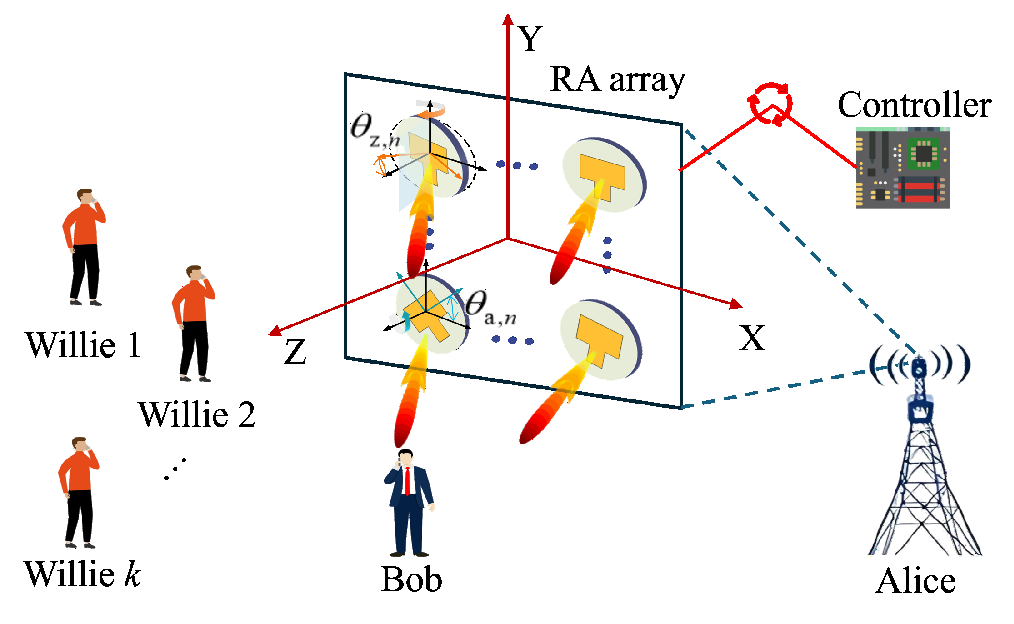} 
	\caption{System model of RA-enabled covert communication.}
	\label{fig:system}
	\vspace{-0.5cm} 
\end{figure}	

		Recently, flexible antenna technologies, including fluid antenna systems (FAS)\cite{FAS}, movable antennas (MA) \cite{movableantenna}, and six-dimensional MA (6DMA)\cite{6DMA}, have garnered significant attention for their ability to dynamically adjust various antenna configurations to improve wireless performance. In particular, rotatable antenna (RA) technology stands out as a simplified yet promising form, emphasizing orientation/boresight rotation flexibility while maintaining fixed antenna positions to offer a cost-effective and compact solution. By mechanically/electronically adjusting the three-dimensional (3D) orientation/boresight direction of individual antennas, RA technology offers enhanced adaptability to dynamic transmission environments and enables the exploitation of additional spatial DoFs. Specifically, the RA-based array can dynamically reconfigure the overall directional gain pattern by independently controlling the rotational angles of each antenna element, thereby actively shaping the overall radiation pattern to enhance array gains in desired directions and improve communication/sensing performance. Up to now, several initial studies have been conducted to demonstrate the significant potential of RA\cite{RAoverview2025,zheng2025rotatableantennaenabledwireless,wumodeling,RAchannelestimation,dai2025rotatableantennaenabledsecurewireless,zhou2025rotatableantennasintegratedsensing,xiethz}. For instance, the authors in  \cite{RAoverview2025} provided a comprehensive overview of RA technology, illustrating RA's capability to flexibly adapt to diverse communication and sensing demands. In particular, the authors in \cite{zheng2025rotatableantennaenabledwireless} and \cite{wumodeling} established a fundamental framework of the RA-enabled communication system, encompassing system modeling, channel characterization, and performance analysis. Furthermore, the authors in \cite{RAchannelestimation} proposed an efficient channel estimation scheme for RA communication systems to improve estimation accuracy. Moreover, studies on RA have been extended into diverse application areas, including secure wireless communications \cite{dai2025rotatableantennaenabledsecurewireless}, integrated sensing and communication (ISAC)\cite{zhou2025rotatableantennasintegratedsensing}, and terahertz (THz) beam squint mitigation \cite{xiethz}. By selectively enhancing or suppressing signal power in desired directions, RA is inherently well suited for scenarios requiring stringent transmission concealment. This makes it particularly advantageous for covert communications, as it can steer signals away from potential wardens (Willies) and reduce the risk of undesired detection.
	
	Motivated by the above, we investigate the application of RA in covert communication systems as shown in Fig.~1. Specifically, an RA-based array is deployed at the base station (Alice), and serves a legitimate user (Bob) covertly in the presence of multiple Willies, who attempt to discern whether or not the transmission exists. Under this framework, we aim to maximize the covert transmission rate, subject to the constraints on the covertness, transmit power, and RA's rotational range. To address the non-convexity of the formulated problem, we propose an efficient alternating optimization (AO) algorithm that leverages the second-order cone programming (SOCP) method and successive convex approximation (SCA) approach to solve the two subproblems iteratively until convergence is reached. Simulation results show that the RA array can significantly enhance the covert transmission rate and outperform other benchmark schemes, especially the fixed-antenna system.
	
	\section{System Model And Problem Formulation}
	
		As shown in Fig. 1, we consider a covert communication system, where Alice wishes to transmit covert information to Bob without being detected by \(K\geq1\) Willies. All Willies and Bob are equipped with a single isotropic antenna. We assume that Alice is equipped with a uniform planar array (UPA) consisting of $N$ directional RAs, which is placed on the $x$-$y$ plane of a 3D Cartesian coordinate system and centered at the origin with $N=N_xN_y$, where $N_x$ and $N_y$ denote the numbers of RAs along $x$- and $y$-axes, respectively. Additionally, the reference positions of the RAs are denoted by \(\mathbf{w}_{n} \in \mathbb{R}^{3 \times 1}\), with \(n \in \{1,2,...,N\}\). Similarly, $\mathbf{q}_k \in \mathbb{R}^{3 \times 1}$, $k \in \{0,1,2,\ldots,K\}$, represent the reference positions of Bob and Willies, where $\mathbf{q}_0$ and $\mathbf{q}_k$, $k\geq1$, are the reference positions of Bob and the $k$-th Willie, respectively. Accordingly, the distance between RA $n$ and Bob/Willie $k$ can be expressed as \(r_{k,n} = \left\|\mathbf{q}_{k} - \mathbf{w}_n \right\|\).

	\subsection{Antenna Boresight Rotation}
	
	The 3D boresight direction of RA $n$ can be characterized by a pointing vector, defined as
		\small
	\begin{equation}
			\vec{\mathbf{f}}_n = [f_{x,n}, f_{y,n}, f_{z,n}]^T \in \mathbb{R}^{3 \times 1},
	\end{equation} \normalsize
	where \(f_{x,n}\), \(f_{y,n}\) and \(f_{z,n}\) are the projections of RA $n$'s pointing vector on the $x$-, $y$- and $z$-axes, respectively. For the boresight rotation of RA $n$, let \(\theta_{z,n} \) denote its zenith angle (i.e., the angle between the boresight direction of RA $n$ and the $z$-axis) and \( \theta_{a,n} \) represent its azimuth angle (i.e., the angle between the projection of the boresight direction of RA $n$ onto the $x$-$y$ plane and the $x$-axis). Accordingly, we have \(f_{x,n} = \sin\theta_{z,n} \cos\theta_{a,n}\), \(f_{y,n} = \sin\theta_{z,n} \sin\theta_{a,n}\) and \(f_{z,n} = \cos\theta_{z,n}\). Furthermore, the zenith angle of each RA should be confined in a specific range \(0 \leq \theta_{\mathrm{z},n} \leq \theta_{\max}, \forall n\), where \(\theta_{\max} \in [0,\frac{\pi}{2}]\) is the maximum zenith angle that each RA is allowed to adjust.

	\subsection{Channel Model}
	We assume that the effective antenna gain of each RA satisfies the following generic directional gain pattern.
		\small
	\begin{equation}
		G_e(\epsilon,\varphi)=
		\begin{cases}
			G_0\cos^{2p}(\epsilon), & \epsilon\in[0,\frac{\pi}{2}),\varphi\in[0,2\pi) \\
			0, & \text{otherwise,}  
		\end{cases}
	\end{equation}\normalsize
	where \((\epsilon,\varphi)\) is a pair of incident angles corresponding to any direction with respect to the RA’s current boresight direction, \(G_0 = 2(2p + 1)\) represents the maximum gain in the boresight direction to meet the law of power conservation, \( p \) is the directivity factor that characterizes the beamwidth of the antenna’s main lobe. Accordingly, the $n$-th RA's directional gain in the direction of Bob/Willie $k$ is given by \(G_0 \cos^{2p}(\epsilon_{k,n})\), where \( \cos(\epsilon_{k,n}) =	\vec{\mathbf{f}}_n^T \vec{\mathbf{q}}_{k,n} \) indicates the projection between \(	\vec{\mathbf{f}}_n\) and \( \vec{\mathbf{q}}_{k,n}\) with \( \vec{\mathbf{q}}_{k,n} = (\mathbf{q}_k - \mathbf{w}_{n}) / r_{k,n}\) denoting the unit direction vector from the $n$-th RA to Bob/Willie $k$.

	Similar to \cite{zheng2025rotatableantennaenabledwireless} and \cite{wumodeling}, we consider the free-space line-of-sight (LoS) propagation model.\footnote{While this work considers a LoS-dominant channel for clarity and tractability, the proposed method is applicable to general multipath environments (e.g., as modeled in \cite{zheng2025rotatableantennaenabledwireless}). This extension involves replacing the LoS channel vectors with multipath channel vectors and re-deriving the corresponding gradient vector and matrix, without altering the overall algorithm design.}. Accordingly, the channels from RA $n$ to Bob/Willie $k$ can be expressed as follows. 
			\small\begin{equation}
	h_{k}(\boldsymbol{\theta}_n) = \sqrt{g_{k}(\boldsymbol{\theta}_n)} e^{-j \frac{2\pi}{\lambda} r_{k,n}},
	\label{eq:channelab}
	\end{equation}\normalsize
	where \(\boldsymbol{\theta}_n=[\theta_{z,n},\theta_{a,n}]^T\) is the rotational angle vector of the $n$-th RA, \(g_{k}(\boldsymbol{\theta}_n) = \frac{S}{4\pi (r_{k,n})^2} G_0 \cos^{2p}(\epsilon_{k,n})\)	denotes the channel power gain between RA $n$ and Bob/Willie $k$, with $S$ being the physical size of each RA element, and $\lambda$ represents the signal wavelength. 

	To explore the performance limit of the proposed RA-enabled covert communication system, we assume that all necessary channel state information (CSI) is available to Alice.\footnote{Although challenging in practice, this assumption serves as a fundamental baseline for evaluating the system's capacity limit and facilitates future robust designs under imperfect CSI. Moreover, obtaining Willie's CSI is technically feasible by detecting the inevitable signal leakage from his superheterodyne receiver (e.g., using the “Ghostbuster” technique \cite{chaman2018ghostbuster}).} According to (\ref{eq:channelab}), the covert transmission rate of Bob is given by 
	\small\begin{equation}
		R_b = \log_2\left(1+\frac{ |\mathbf{w}^H{\mathbf{h}_{0}}(\boldsymbol{\Theta})|^2}{\sigma_b^2}\right),
	\end{equation}\normalsize
	where \( \mathbf{w}^H \in \mathbb{C}^{1 \times N} \) is the transmit beamforming vector, \(\mathbf{h}_{0}(\boldsymbol{\Theta})\stackrel{\triangle}{=}[h_{0}(\boldsymbol{\theta}_1),h_{0}(\boldsymbol{\theta}_2),...,h_{0}(\boldsymbol{\theta}_N)]^T \) is the channel vector from Alice to Bob, \(\boldsymbol{\Theta}\stackrel{\triangle}{=}[\boldsymbol{\theta}_1,\boldsymbol{\theta}_2,...,\boldsymbol{\theta}_N]\in \mathbb{R}^{2 \times N}\) denotes the antenna rotation matrix, and \(\sigma_b^2\) is the additive white Gaussian noise (AWGN) power at Bob.

	\subsection{Binary Hypothesis Testing at Willie}
	For covert transmission, we assume that each Willie performs independent energy detection to determine whether the transmission from Alice to Bob exists or not. Considering the identical noise uncertainty \(\rho\in[1,\infty)\) at each Willie\cite{Noise}, we define the null hypothesis \(\mathcal{H}_0\) as the case where Alice remains silent, and the alternative hypothesis \(\mathcal{H}_1\) as the case where Alice transmits information to Bob. As a result, the observed signal at Willie $k$ can be expressed as
		\small
	\begin{equation}
	 y_{k}[l] = 
	\begin{cases}
		n_{k}[l], & \mathcal{H}_0 \\
		\mathbf{w}^H \mathbf{h}_{k}(\boldsymbol{\Theta}) s_a[l] + n_{k}[l], & \mathcal{H}_1,
	\end{cases}
	\end{equation}\normalsize
where \(\mathbf{h}_{k}(\boldsymbol{\Theta})\stackrel{\triangle}{=}[h_{k}(\boldsymbol{\theta}_1),h_{k}(\boldsymbol{\theta}_2),...,h_{k}(\boldsymbol{\theta}_N)]^T\) is the channel vector from Alice to Willie $k$, and \(s_a[l]\) is the signal transmitted by Alice, which satisfies \( \mathbb{E}({|s_a[l]|^2})=1 \), with \(l\in\{1,2,\ldots,L\}\) denoting the index of channel use, and \(n_{k}[l] \sim \mathcal{CN}(0, \sigma_w^2)\) is the AWGN at Willie $k$ with  \({\sigma}^2_w\) being the noise power. Each Willie makes a binary decision by comparing its average power \(T_w\) of the received signals with a preset power detection threshold \(\Gamma\). The decision rule is given by 	\small\(T_w = \frac{1}{L} \sum_{l = 1}^{L} \left|y_{k}[l]\right|^2\underset{D_0}{\overset{D_1}{\gtrless}}
\Gamma\)\normalsize, where \(\mathcal{D}_0\) and \(\mathcal{D}_1\) represent the decision results supporting \(\mathcal{H}_0\) and \(\mathcal{H}_1\), respectively.
	
	Under the assumption that the priori probability of hypotheses \(\mathcal{H}_0\) and \(\mathcal{H}_1\) being true is equal (i.e., both are 0.5), then the conditional detection error probability (DEP) of Willie $k$ is given by \cite{Noise}
	\small\begin{equation}
    	\begin{aligned}
		\xi_k &= P_{FA,k} + P_{MD,k} \\
			  &= 1 - \Pr((\Gamma - v_k) < \sigma^2_w < \Gamma),
	\end{aligned}
	\end{equation}\normalsize
	where \(v_{k}=|\mathbf{w}^H \mathbf{h}_{k}(\boldsymbol{\Theta})|^2\), and \(P_{FA,k}=\Pr(\mathcal{D}_1|\mathcal{H}_0)\) and \(P_{MD,k}=\Pr(\mathcal{D}_0|\mathcal{H}_1)\) denote the false alarm and miss detection probabilities of each Willie, respectively.

	We assume that each Willie adopts the optimal decision rule as specified in \cite{Noise}, and thus the minimum DEP for each Willie can be analytically obtained as
	\small\begin{equation}
    \xi_k^* = 1 - \frac{1}{2\ln\rho}\ln\left(1 + \frac{\rho v_k}{{\tilde{\sigma}^2_w}}\right), \\
	\label{eq:DEP}
	\end{equation}\normalsize
where $\tilde{\sigma}^2_w$ is the nominal noise power without noise uncertainty \cite{Noise}.

To ensure the covertness constraint \(\xi_k^*\geq1-\delta\), where \(\delta\) denotes the maximum allowable detection probability (i.e., the covertness tolerance level), the condition in (\ref{eq:DEP}) can be equivalently reformulated as the following constraint on the received signal power at each Willie:
	\small	\begin{equation}
	\quad |\mathbf{w}^H \mathbf{h}_{k}(\boldsymbol{\Theta})|^2 \leq \eta,  \quad\forall k = 1, \ldots, K,
	\end{equation}\normalsize
	where \(\eta = \min\left(\tilde{\sigma}_w^2 (\rho - \frac{1}{\rho}), \frac{(\rho^{2\delta}-1){\tilde\sigma}_w^2}{\rho}\right)\) is the maximum allowable received signal power threshold at each Willie to satisfy the covertness requirement.
	\subsection{Problem Formulation}
	To maximize the covert transmission rate of Bob subject to the constraints on covertness requirement, transmit power, and RA's rotational range, the optimization problem can be formulated as
	\begin{subequations}
		\label{eq:problem_first}
	\small\begin{align}
		\max_{\mathbf{w},\boldsymbol{\Theta}} &\ \log_2\left(1 + \frac{|\mathbf{w}^H \mathbf{h}_{0}(\boldsymbol{\Theta})|^2}{\sigma_b^2}\right) \label{eq:obj}\\
		\text{s.t. } &\ |\mathbf{w}^H \mathbf{h}_{k}(\boldsymbol{\Theta})|^2 \leq \eta, \quad\forall k = 1,\ldots,K, \label{eq:con1}\\
		&\ \|\mathbf{w}\|^2 \leq P_{\max}, \label{eq:con2}\\
		&\ 0 \leq \theta_{z,n} \leq \theta_{\max}, \quad n = 1,2,\ldots,N, \label{eq:con3}\\
		&\ 0 \leq \theta_{a,n} \leq 2\pi, \quad n = 1,2,\ldots,N, \label{eq:con4}
	\end{align}\normalsize
	\end{subequations}
where \(P_{\max}\) is the maximum power budget of Alice. Constraints (\ref{eq:con1}) and (\ref{eq:con2}) represent the covertness and transmit power constraints, respectively. The constraints on RA's rotational range are given in  (\ref{eq:con3}) and  (\ref{eq:con4}). Note that problem (\ref{eq:problem_first}) is difficult to solve directly due to the non-concavity of the objective function as well as the coupling between the transmit beamforming vector \( \mathbf{w} \) and RA's rotational angles \( \boldsymbol{\Theta} \). To efficiently tackle the problem, we propose an AO algorithm to alternately optimize the transmit beamforming vector and RA's rotational angles.
	\section{Proposed Algorithm}
	In this section, we employ the AO algorithm to solve the two subproblems derived from problem (\ref{eq:problem_first}) alternately until convergence is reached.

	\subsection{Subproblem 1: Transmit Beamforming Optimization}
	
With fixed rotational angles \( \boldsymbol{\Theta} \), the problem (\ref{eq:problem_first}) can be reformulated as
	\begin{subequations}
		\label{eq:problem_second}
	\small\begin{align}
		\max_{\mathbf{w}} &\ |\mathbf{w}^H \mathbf{h}_{0}|^2 \label{eq:subproblem_w}\\		
		\text{s.t. } &\ |\mathbf{w}^H \mathbf{h}_{k}|^2 \leq \eta, \quad\forall k = 1, \ldots, K, \label{eq:sub1con1}\\
		&\ ||\mathbf{w}\|^2 \leq P_{\max} \label{eq:sub1con2}.
	\end{align}
	\end{subequations}\normalsize
	It is worth noting that, although the constraints in problem (\ref{eq:problem_second}) are a convex set with respect to \( \mathbf{w} \), the non-concave objective function renders the overall problem non-convex. Nevertheless, we can observe that if \( \mathbf{w} \) satisfies the above two constraints, so does \( e^{j\theta}\mathbf{w}\) for any arbitrary $\theta$, and the value of the objective function is maintained. Therefore, we assume that \(\mathbf{w}^H \mathbf{h}_{0}\) is a real value. Moreover, by introducing a slack optimization variable $t$\cite{Exploit}, the problem (\ref{eq:problem_second}) can be rewritten as
	\begin{subequations}
			\label{eq:problem_sub1solve}
	\small\begin{align}
		\max_{\mathbf{w},t} &\ t \label{eq:subproblem_w}\\		
		\text{s.t. } &\  \mathrm{Re}(\mathbf{w}^H \mathbf{h}_{0}) \geq t,\\
		&\ \mathrm{Im}(\mathbf{w}^H \mathbf{h}_{0}) = 0,\\
		&\ |\mathbf{w}^H \mathbf{h}_{k}|^2 \leq \eta, \quad\forall k = 1, \ldots, K, \\
		&\ ||\mathbf{w}\|^2 \leq P_{\max}.
	\end{align}\normalsize
\end{subequations}
It is evident that the above problem is a second-order cone programming \cite{boyd2004convex}, which can be efficiently solved by standard numerical optimization software.
	\subsection{Subproblem 2: Antenna Rotation Optimization}
	With fixed transmit beamforming \( \mathbf{w} \), problem (\ref{eq:problem_first}) can be reformulated as 
	\begin{subequations}
			\label{eq:sub2}
		\small	\begin{align}
		\max_{\boldsymbol{\Theta}}  &\ |\mathbf{w}^H \mathbf{h}_{0}(\boldsymbol{\Theta})|^2, 	\label{eq:sub2obj}\\
			\text{s.t} &\ \eqref{eq:con1}, \eqref{eq:con3}, \eqref{eq:con4}.
		\end{align}\normalsize
	\end{subequations}
However, this subproblem remains challenging to solve due to the non-concavity of the objective function and the non-convexity of the constraint \eqref{eq:con1} with respect to RA's rotational angles. To tackle this problem, we first rewrite the objective function as
		\small\begin{align}
		\left| \mathbf{w}^H \mathbf{h}_{0}(\boldsymbol{\Theta}) \right|^2 
		&= \sum_{n = 1}^N \sum_{m = 1}^N |w_n w_m| c_{nm} z_{nm},
		\label{eq:objective_rewritten}
	\end{align}\normalsize
where \(z_{nm}=\cos^p\left(\epsilon(\boldsymbol{\theta}_n)\right) \cos^p(\epsilon(\boldsymbol{\theta}_m))\cos(\psi_{nm}^i)\), \(\psi_{nm}^i=\psi_{nm} - (\angle w_m - \angle w_n)\), \( c_{nm} = \frac{SG_0}{4\pi r_{b,n} r_{b,m}} \), and \(\psi_{nm} = \frac{2\pi}{\lambda} (r_{b,m} - r_{b,n})\). Note that the constant term (i.e.,  \(\sum_{n = 1}^N \sum_{m = 1}^N |w_n w_m| c_{nm}\)) does not affect the concavity of the objective function. Therefore, we only need to apply a second-order Taylor (SOT) expansion on \(z_{nm}\) to derive a tractable surrogate.  In particular, we first obtain a concave lower bound for \(\cos\left(\epsilon(\boldsymbol{\theta}_n)\right)\) at the current initial point \(\boldsymbol{\theta}_n^i\) via SOT expansion:
\small	\begin{align}
	\cos\left(\epsilon(\boldsymbol{\theta}_n)\right) \approx \cos(\epsilon(\boldsymbol{\theta}_n^i)) + \mathbf{d}_n^T \Delta \boldsymbol{\theta}_n + \frac{1}{2} \Delta \boldsymbol{\theta}_n^T \mathbf{H}_n \Delta \boldsymbol{\theta}_n, \label{eq:cos_eps_lower}
\end{align}\normalsize

	\small	\twocolumn[
\begin{center}
	\begin{align}
		\mathbf{A}_n=& -p \sum_{m=1}^N |w_n w_m| c_{nm} \left(\cos^{p-1}(\epsilon(\boldsymbol{\theta}_n^i)) \cos^p(\epsilon(\boldsymbol{\theta}_m^i))+\cos^{p}(\epsilon(\boldsymbol{\theta}_n^i)) \cos^{p-1}(\epsilon(\boldsymbol{\theta}_m^i))\right) \cos(\psi_{nm}^i) \mathbf{I}_2, n=1,2,...,N. \label{eq:A_diag}
	\end{align}	
\end{center}
\vspace{-1.5mm} 
\noindent\rule{\textwidth}{0.4pt} 
]	\normalsize
\noindent where \(\Delta \boldsymbol{\theta}_n = [\Delta \theta_{z,n}, \Delta \theta_{a,n}]^T\) represents the deviation from the current point, with \(\Delta \theta_{z,n} = \theta_{z,n} - \theta_{z,n}^i\) and \(\Delta \theta_{a,n} = \theta_{a,n} - \theta_{a,n}^i\). Defining \(\alpha=\frac{\partial \cos(\epsilon(\boldsymbol{\theta}_n))}{\partial \epsilon(\boldsymbol{\theta}_n)}\), then the gradient vector \(\mathbf{d}_n\) and the Hessian matrix \(\mathbf{H}_n\) can be represented as
	\small\begin{align}
	   \mathbf{d}_n = \begin{bmatrix}
			\frac{\alpha\partial\epsilon(\boldsymbol{\theta}_n)}{\partial \theta_{z,n}} \\
			\frac{\alpha\partial\epsilon(\boldsymbol{\theta}_n)}{\partial \theta_{a,n}}
		\end{bmatrix},
		\mathbf{H}_n = \begin{bmatrix} \frac{\alpha^2\partial^2 \epsilon(\boldsymbol{\theta}_n)}{\partial^2 \theta_{z,n}} & \frac{\alpha^2\partial^2 \epsilon(\boldsymbol{\theta}_n)}{\partial \theta_{z,n} \partial \theta_{a,n}} \\ \frac{\alpha^2\partial^2 \epsilon(\boldsymbol{\theta}_n)}{\partial \theta_{z,n} \partial \theta_{a,n}} & \frac{\alpha^2\partial^2 \epsilon(\boldsymbol{\theta}_n)}{\partial^2 \theta_{a,n}} \end{bmatrix}.
	\end{align}\normalsize

Since  \(\cos\left(\epsilon(\boldsymbol{\theta}_n)\right)\) is smooth, its second-order derivatives are bounded by 1 in magnitude, implying that the Hessian matrix \(\mathbf{H}_n\) satisfies \(-\mathbf{I}_2\preceq\mathbf{H}_n\preceq\mathbf{I}_2\), where \(\mathbf{I}_2\) is the \(2\times2\) identity matrix. This enables the construction of a concave lower bound by substituting \(\mathbf{H}_n\) with \(-\mathbf{I}_2\), which yields the following expression,
    \small\begin{align}
    		\cos\left(\epsilon(\boldsymbol{\theta}_n)\right) \geq 	\cos(\epsilon(\boldsymbol{\theta}_n^i)) + \mathbf{d}_n^T \Delta \boldsymbol{\theta}_n - \frac{1}{2} \Delta \boldsymbol{\theta}_n^T \mathbf{I}_2 \Delta \boldsymbol{\theta}_n. \label{eq:cos_eps_lower_bound}
    \end{align}\normalsize

Next, the above approximation is extend from \(\cos\left(\epsilon(\boldsymbol{\theta}_n)\right)\) to \(\cos^p(\epsilon(\boldsymbol{\theta}_n))\). Assuming \(\cos\left(\epsilon(\boldsymbol{\theta}_n)\right) \geq 0\) (a condition can be enforced by appropriately selecting the initial point), the lower bound of \(\cos^p(\epsilon(\boldsymbol{\theta}_n))\) becomes
	\small\begin{align}
	\cos^p(\epsilon(\boldsymbol{\theta}_n)) 
	&\geq \cos^p(\epsilon(\boldsymbol{\theta}_n^i)) + p \cos^{p - 1}(\epsilon(\boldsymbol{\theta}_n^i)) \notag \\
	&\quad\times\left( \mathbf{d}_n^T \Delta \boldsymbol{\theta}_n - \frac{1}{2} \Delta \boldsymbol{\theta}_n^T \mathbf{I}_2 \Delta \boldsymbol{\theta}_n \right). 
	\end{align}\normalsize

A similar expression holds for \(\cos^p(\epsilon(\boldsymbol{\theta}_m))\) in the following, 
	\small\begin{align}
		\cos^p(\epsilon(\boldsymbol{\theta}_m)) 
		&\geq \cos^p(\epsilon(\boldsymbol{\theta}_m^i)) + p \cos^{p-1}(\epsilon(\boldsymbol{\theta}_m^i)) \notag \\
		&\quad\times\left( \mathbf{d}_m^T \Delta \boldsymbol{\theta}_m - \frac{1}{2} \Delta \boldsymbol{\theta}_m^T \mathbf{I}_2 \Delta \boldsymbol{\theta}_m \right).
	\end{align}\normalsize

Thus, by substituting the above derived lower bounds of \(\cos^p(\epsilon(\boldsymbol{\theta}_n))\) and \(\cos^p(\epsilon(\boldsymbol{\theta}_m))\) into the expression of \( z_{nm} \), the lower bound of the objective function can be expressed as
	\small\begin{align}
		\left| \mathbf{w}^H \mathbf{h}_{0}(\boldsymbol{\Theta}) \right|^2 &\geq c + \mathbf{b}^T \Delta \boldsymbol{\Theta} + \frac{1}{2} \Delta \boldsymbol{\Theta}^T \mathbf{A} \Delta \boldsymbol{\Theta}, \label{eq:objective_lower_bound}
	\end{align}\normalsize
where \(\Delta \boldsymbol{\Theta} = [\Delta \boldsymbol{\theta}_1^T, \ldots, \Delta \boldsymbol{\theta}_N^T]^T \in \mathbb{C}^{2N \times 1}\), the coefficients \(c\), \(\mathbf{b}\), and \(\mathbf{A}\) are derived from the constant, linear, and quadratic terms of the lower bound of \(z_{nm}\), respectively. In particular, the constant coefficient $c$ is \(\sum_{n=1}^N \sum_{m=1}^N |w_n w_m| c_{nm} \cos^p(\epsilon(\boldsymbol{\theta}_n^i)) \cos^p(\epsilon(\boldsymbol{\theta}_m^i)) \cos(\psi_{nm}^i)\), and the $n$-th element of the linear coefficient \(\mathbf{b}\) is expressed as  \(\mathbf{b}_n= \sum_{m=1}^N |w_n w_m| c_{nm} p \cos^{p-1}(\epsilon(\boldsymbol{\theta}_n^i)) \cos^p(\epsilon(\boldsymbol{\theta}_m^i))\times\\\cos(\psi_{nm}^i) \mathbf{d}_n\).

	As for the Hessian matrix \(\mathbf{A} \in \mathbb{R}^{2N \times 2N}\), we neglect the cross-coupling effects among different antennas, and thus only retain the diagonal blocks of \(\mathbf{A}\) to reduce computational complexity, as defined in (\ref{eq:A_diag}) shown at the top of this page. Each diagonal block \(\mathbf{A}_n\) is scaled by non-positive  second-order derivatives and non-negative weights, thereby ensuring that all eigenvalues of \(\mathbf{A}\) are non-positive. This leads to the conclusion that  \(\mathbf{A} \preceq \mathbf{0}\), which guarantees the concavity of the quadratic surrogate in (\ref{eq:objective_lower_bound}).

	Regarding the constraint (\ref{eq:con1}), we adopt a similar approximation strategy by replacing \(\mathbf{H}_n\) with \(\mathbf{I}_2\) to construct its upper bound. Consequently, \(\tilde{c}_k\), \(\tilde{\mathbf{b}}_k\), and \(\tilde{\mathbf{A}}_k\) can be derived in a similar way, with \(\tilde{\mathbf{A}}_k \succeq \mathbf{0}\) denoting a positive semi-definite matrix. Then, the approximated constraint function can be reformulated as
\small\begin{align}
	|\mathbf{w}^H \mathbf{h}_{k}(\boldsymbol{\Theta})|^2 \leq \tilde{c}_k + \tilde{\mathbf{b}}_k^T \Delta \boldsymbol{\Theta} + \frac{1}{2} \Delta \boldsymbol{\Theta}^T \tilde{\mathbf{A}}_k \Delta \boldsymbol{\Theta}.\label{eq:constraint_upper_bound}
\end{align}\normalsize

Following the above analysis, subproblem 2 can be transformed into
\begin{subequations}
	\label{eq:sub2_final}
\small\begin{align}
	\max_{\Delta  \boldsymbol{\Theta}} \quad & c + \mathbf{b}^T \Delta  \boldsymbol{\Theta} + \frac{1}{2} \Delta  \boldsymbol{\Theta}^T \mathbf{A} \Delta  \boldsymbol{\Theta} \\
	\text{s.t.} \quad & \tilde{c}_k + \tilde{\mathbf{b}}_k^T \Delta  \boldsymbol{\Theta} + \frac{1}{2} \Delta  \boldsymbol{\Theta}^T \tilde{\mathbf{A}}_k \Delta  \boldsymbol{\Theta} \leq \eta, \quad\forall k = 1, \ldots, K, \\
	 &\eqref{eq:con3},\eqref{eq:con4}.
\end{align}\normalsize
\end{subequations}
The above problem \eqref{eq:sub2_final} is a convex optimization problem, which can be efficiently solved by the CVX tool \cite{boyd2004convex}.

	\subsection{Overall Algorithm}
		\vspace{-0.35cm}
		\begin{algorithm}
		\caption{Proposed AO Algorithm}
		\label{alg:alternating}
		\begin{algorithmic}[1]
			\State \textbf{Initialize} \( i = 0 \) and \( \boldsymbol{\Theta}^{0}=\mathbf{0}_{2 \times N} \).
			\Repeat
			\State \( i = i + 1 \).
			\State Obtain \( \mathbf{w}^{i + 1} \) by solving problem \eqref{eq:problem_sub1solve} with given \( \boldsymbol{\Theta}^{i} \).
			\State Obtain \( \boldsymbol{\Theta}^{i + 1} \) by solving problem \eqref{eq:sub2_final} with given \( \mathbf{w}^{i + 1} \).
			\Until The fractional increase of the covert transmission rate $R_b$ in (4) is below a threshold $\xi > 0$ or the iteration number $i$ reaches the pre-designed number of iterations $I$.
			\State \Return \( \mathbf{w}^{opt} = \mathbf{w}^{i}\) and \( \boldsymbol{\Theta}^{opt} = \boldsymbol{\Theta}^{i}\).
		\end{algorithmic}
	\end{algorithm}
		\vspace{-0.15cm}
The overall AO algorithm to solve problem \eqref{eq:problem_first} is presented in Algorithm 1. Since the transmit beamforming vector $\mathbf{w}$ is obtained via problem \eqref{eq:problem_sub1solve}, and the rotational angle matrix $\boldsymbol{\Theta}$ is updated by solving a convex problem that further improves or maintains the objective, each iteration ensures a non-decreasing objective value of problem \eqref{eq:problem_first}. Given that the objective is upper-bounded by a finite value, the convergence of Algorithm 1 is thus guaranteed. In terms of computational complexity, the cost of updating the transmit beamforming vector is \small\(\mathcal{O} \left(N^{3}\sqrt{K} \right)\)\normalsize, while solving problem \eqref{eq:sub2_final} incurs a complexity of \small\(\mathcal{O}\left((2N + K)^{3.5} \ln\left(1/{\varepsilon}_1\right)\right)\)\normalsize  per iteration, where \(\varepsilon_1\) represents the accuracy threshold for convergence in subproblem 2. Hence, the complexity of Algorithm 1 is \small\(\mathcal{O}\left(I\left(N^{3}\sqrt{K} +(2N + K)^{3.5} \ln(1/{\varepsilon_1})\right)\right)\)\normalsize, with $I$ denoting the number of iterations required for convergence.
	\vspace{-0.35cm} 

	\section{Simulation Results}

		\begin{figure*}[!t]
	\centering
	\begin{minipage}[t]{0.32\textwidth}
		\centering
		\includegraphics[width=\linewidth]{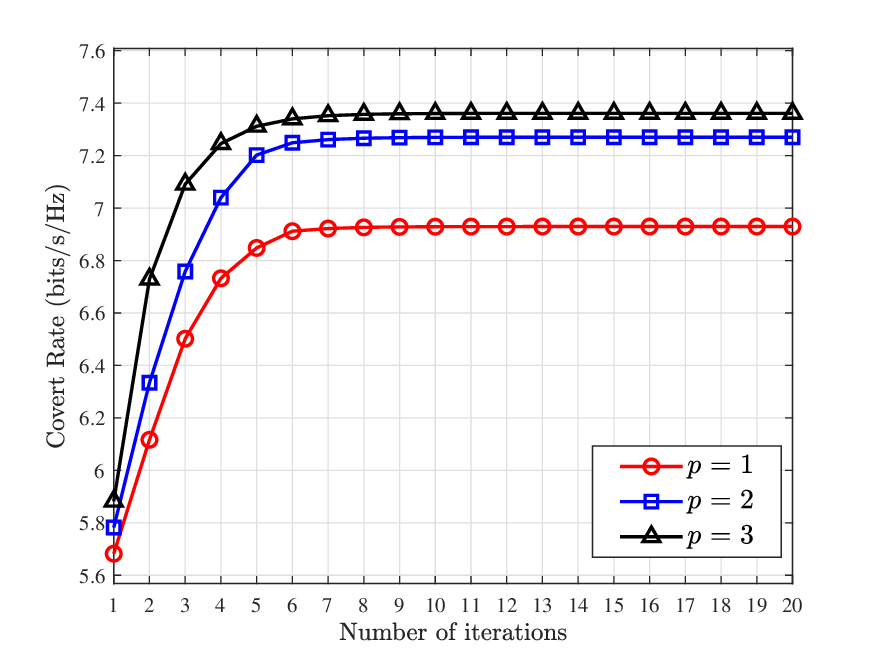} 
		\vspace{-0.8cm} 
		\caption{Convergence behavior of the proposed AO algorithm.}
		\label{fig2}
	\end{minipage}
	\hfill 
	\begin{minipage}[t]{0.32\textwidth}
		\centering
		\includegraphics[width=\linewidth]{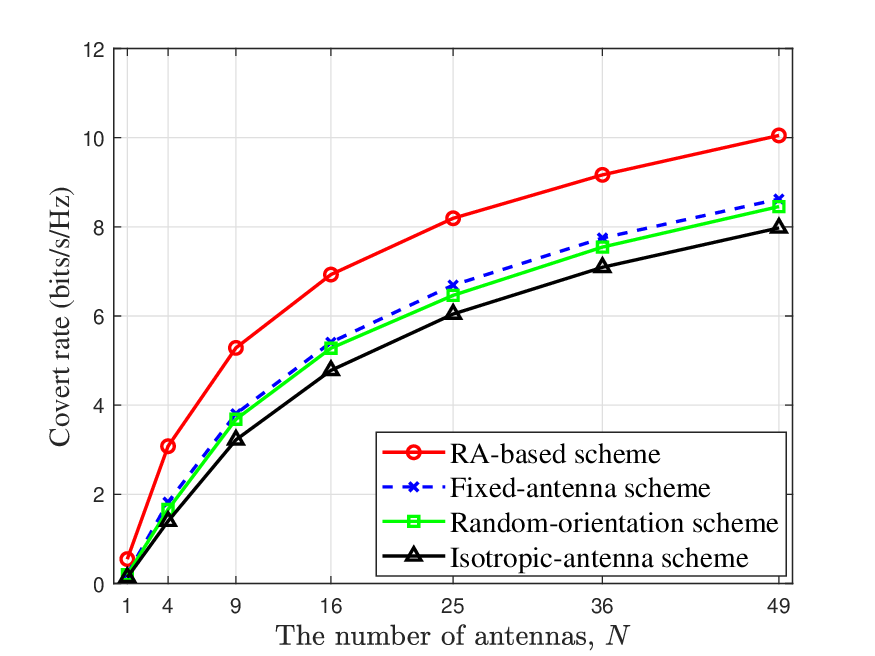} 
		\vspace{-0.8cm} 
		\caption{Covert rate $R_b$ versus the numbers of antennas $N$.}
		\label{fig3}
	\end{minipage}
	\hfill 
	\begin{minipage}[t]{0.32\textwidth}
		\centering
		\includegraphics[width=\linewidth]{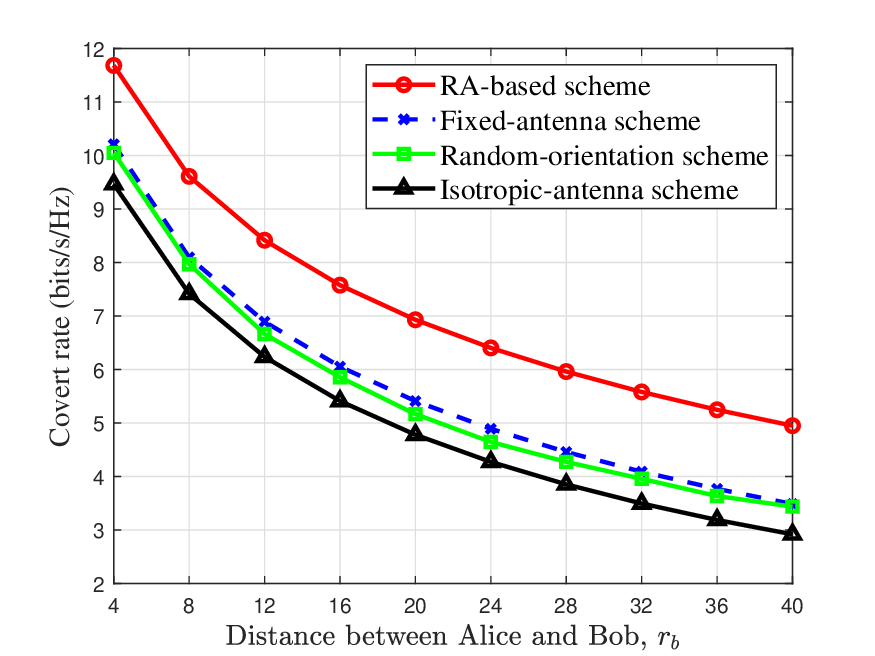} 
		\vspace{-0.8cm} 
		\caption{Covert rate $R_b$ versus the distance between Alice and Bob.}
		\label{fig4}
	\end{minipage}
	\vspace{-0.4cm} 
\end{figure*}

	In this section, we present simulation results to evaluate the performance of the proposed RA-enabled covert communication system. We compare it with three benchmark schemes: (1) the fixed-antenna scheme, where the orientations of all RAs are fixed at their reference orientations, i.e., \(\boldsymbol{\Theta}=\mathbf{0}_{2 \times N}\); (2) the random-orientation scheme, in which each antenna's orientation is randomly generated within the rotational range; (3) the isotropic-antenna scheme, in which the directional gain of each antenna is set to $p=0$. Unless otherwise stated, we set the simulation parameters as follows: the wavelength is $\lambda = 0.125$ meter (m), the noise power $\tilde{\sigma}^2_w=\sigma^2_b=-90 $ dBm, the covertness tolerance level is $\delta=0.01$, the antenna directivity exponent is $p=1$, the maximum adjustable angle of each RA is \(\theta_{\max}=\frac{\pi}{6}\), and the noise uncertainty at each Willie is $\rho=3$ dB\cite{Noise}. We also assume Bob and $K=2$ Willies are located at coordinates \([r_b\cos\phi_b,0,r_b\sin\phi_b]\), \([r_{w1}\cos\phi_{w1},0,r_{w1}\sin\phi_{w1}]\), and \([r_{w2}\cos\phi_{w2},0,r_{w2}\sin\phi_{w2}]\), respectively,  where $r_{w1}=r_{w2}=30$ m, and the angular positions are fixed as \(\phi_b=\frac{\pi}{3}\), \(\phi_{w1}=\frac{\pi}{4}\), and \(\phi_{w2}=\frac{3\pi}{4}\). The performance results below are averaged over 100 independent random realizations to eliminate fluctuations.

	Fig. 2 illustrates the convergence behavior of the proposed AO algorithm. As expected, the covert rate monotonically increases and stabilizes within approximately 8 iterations across all considered antenna directivity factors $p$. Furthermore, the covert rate increases with $p$, as the larger $p$ provides a higher directional gain and narrower beamwidth, thereby enabling the RA system to more effectively concentrate radiation power toward the desired direction.
	
	In Fig. 3, we present the covert transmission rate $R_b$ against the number of antennas $N$, where $P_{\max}=30$ dBm and $r_b=20$~m. It can be observed that the RA-based system consistently outperforms other benchmark schemes. This is attributed to the RA's ability to dynamically adjust the array directional gain pattern through optimization of the antenna rotational angles, thereby focusing the radiation power toward desired directions and enhancing covert transmission rate. In contrast, the fixed-antenna scheme focuses radiation power toward a fixed direction and cannot adapt the boresight direction to serve the legitimate user effectively. Similarly, the random-orientation scheme does not strategically design the antenna orientation/boresight directions, yielding inferior array gain in the legitimate user's direction compared to the proposed RA-based scheme. These results validate the effectiveness of the proposed RA-based scheme in flexibly reconfiguring the array directional gain pattern to enhance covert communication performance.
	
	In Fig. 4, we plot the covert transmission rate $R_b$ versus the distance between Alice and Bob with $N$ = 16, and $P_{\max}=30$~dBm. It is observed that as the distance increases, the covert rates of all schemes decrease subsequently. This is because the increasing distance leads to a higher path loss, which in turn reduces the received signal power at Bob. Nevertheless, the RA-enabled system consistently achieves the highest covert rate among all the schemes, even at larger transmission distances. This is attributed to the fact that the RA-enabled system can dynamically adjust each antenna's orientation to obtain a desired directional gain pattern, thereby mitigating the impact of path loss more effectively than other baseline schemes.

	\section{Conclusions}
	In this letter, we proposed a new RA-enabled covert communication system, where each antenna at Alice can flexibly adjust its boresight direction to dynamically reconfigure the overall directional gain pattern. By jointly optimizing the transmit beamforming and rotational angles of each RA, we developed an efficient AO algorithm to maximize the covert transmission rate of Bob, while satisfying the covertness constraint at each Willie. Simulation results demonstrated that the proposed RA-enabled covert communication system achieves a significantly higher covert rate than various benchmark schemes, even under the adverse conditions with multiple Willies. This underscores its strong potential for undetectable communication in surveillance-intensive environments.
	
		\vspace{-0.37cm} 
	\bibliographystyle{IEEEtran} 
	\bibliography{references} 

\begin{thebibliography}{10}
\providecommand{\url}[1]{#1}
\csname url@samestyle\endcsname
\providecommand{\newblock}{\relax}
\providecommand{\bibinfo}[2]{#2}
\providecommand{\BIBentrySTDinterwordspacing}{\spaceskip=0pt\relax}
\providecommand{\BIBentryALTinterwordstretchfactor}{4}
\providecommand{\BIBentryALTinterwordspacing}{\spaceskip=\fontdimen2\font plus
\BIBentryALTinterwordstretchfactor\fontdimen3\font minus
  \fontdimen4\font\relax}
\providecommand{\BIBforeignlanguage}[2]{{%
\expandafter\ifx\csname l@#1\endcsname\relax
\typeout{** WARNING: IEEEtran.bst: No hyphenation pattern has been}%
\typeout{** loaded for the language `#1'. Using the pattern for}%
\typeout{** the default language instead.}%
\else
\language=\csname l@#1\endcsname
\fi
#2}}
\providecommand{\BIBdecl}{\relax}
\BIBdecl

\bibitem{Bash}
B.~A. Bash, D.~Goeckel, and D.~Towsley, ``Limits of reliable communication with
  low probability of detection on {AWGN} channels,'' \emph{IEEE J. Sel. Areas
  Commun.}, vol.~31, no.~9, pp. 1921--1930, Sept. 2013.

\bibitem{Noise}
B.~He, S.~Yan, X.~Zhou, and V.~K.~N. Lau, ``On covert communication with noise
  uncertainty,'' \emph{IEEE Commun. Lett.}, vol.~21, no.~4, pp. 941--944, Jan.
  2017.

\bibitem{channel}
J.~Wang, W.~Tang, Q.~Zhu, X.~Li, H.~Rao, and S.~Li, ``Covert communication with
  the help of relay and channel uncertainty,'' \emph{IEEE Wireless Commun.
  Lett.}, vol.~8, no.~1, pp. 317--320, Sept. 2019.

\bibitem{AN}
R.~Soltani, D.~Goeckel, D.~Towsley, B.~A. Bash, and S.~Guha, ``Covert wireless
  communication with artificial noise generation,'' \emph{IEEE Trans. Wireless
  Commun.}, vol.~17, no.~11, pp. 7252--7267, Nov. 2018.

\bibitem{FAS}
K.-K. Wong, A.~Shojaeifard, K.-F. Tong, and Y.~Zhang, ``Fluid antenna
  systems,'' \emph{IEEE Trans. Wireless Commun.}, vol.~20, no.~3, pp.
  1950--1962, Mar. 2021.

\bibitem{movableantenna}
L.~Zhu, W.~Ma, and R.~Zhang, ``Movable antennas for wireless communication:
  Opportunities and challenges,'' \emph{IEEE Commun. Mag.}, vol.~62, no.~6, pp.
  114--120, Jun. 2024.

\bibitem{6DMA}
X.~Shao, Q.~Jiang, and R.~Zhang, ``6{D} movable antenna based on user
  distribution: Modeling and optimization,'' \emph{IEEE Trans. Wireless
  Commun.}, vol.~24, no.~1, pp. 355--370, Jan. 2025.

\bibitem{RAoverview2025}
B.~Zheng, T.~Ma, C.~You, J.~Tang, R.~Schober, and R.~Zhang, ``Rotatable antenna
  enabled wireless communication and sensing: Opportunities and challenges,''
  \emph{{IEEE} Wireless Commun.}, 2025, {E}arly Access.

\bibitem{zheng2025rotatableantennaenabledwireless}
B.~Zheng, Q.~Wu, and R.~Zhang, ``Rotatable antenna enabled wireless
  communication: Modeling and optimization,'' \emph{{IEEE} Trans. Commun.},
  2026, {E}arly Access.

\bibitem{wumodeling}
Q.~Wu, B.~Zheng, T.~Ma, and R.~Zhang, ``Modeling and optimization for rotatable
  antenna enabled wireless communication,'' in \emph{Proc. IEEE Int. Conf.
  Commun. (ICC)}, Montreal, Canada, 2025, pp. 1055--1060.

\bibitem{RAchannelestimation}
X.~Xiong, B.~Zheng, W.~Wu, X.~Shao, L.~Dai, M.-M. Zhao, and J.~Tang,
  ``Efficient channel estimation for rotatable antenna-enabled wireless
  communication,'' \emph{{IEEE} Wireless Commun. Lett.}, vol.~14, no.~11, pp.
  3719--3723, Nov. 2025.

\bibitem{dai2025rotatableantennaenabledsecurewireless}
L.~Dai, B.~Zheng, Q.~Wu, C.~You, R.~Schober, and R.~Zhang, ``Rotatable
  antenna-enabled secure wireless communication,'' \emph{{IEEE} Wireless
  Commun. Lett.}, vol.~14, no.~11, pp. 3440--3444, Nov. 2025.

\bibitem{zhou2025rotatableantennasintegratedsensing}
C.~Zhou, C.~You, B.~Zheng, X.~Shao, and R.~Zhang, ``Rotatable antennas for
  integrated sensing and communications,'' \emph{{IEEE} Wireless Commun.
  Lett.}, vol.~14, no.~9, pp. 2838--2842, Sept. 2025.

\bibitem{xiethz}
Y.~Xie, W.~Mei, D.~Wang, B.~Ning, Z.~Chen, J.~Fang, and W.~Guo, ``{TH}z beam
  squint mitigation via 3{D} rotatable antennas,'' in \emph{Proc. IEEE Int.
  Conf. Commun. Workshops}, Montreal, Canada, 2025, pp. 26--31.

\bibitem{chaman2018ghostbuster}
A.~Chaman, J.~Wang, J.~Sun, H.~Hassanieh, and R.~R. Choudhury, ``Ghostbuster:
  Detecting the presence of hidden eavesdroppers,'' in \emph{Proc. 24th Annu.
  Int. Conf. Mobile Comput. Netw. MobiCom}, 2018, pp. 337--351.

\bibitem{Exploit}
R.~Zhang and Y.-C. Liang, ``Exploiting multi-antennas for opportunistic
  spectrum sharing in cognitive radio networks,'' \emph{IEEE J. Sel. Topics
  Signal Process.}, vol.~2, no.~1, pp. 88--102, Feb. 2008.

\bibitem{boyd2004convex}
S.~Boyd and L.~Vandenberghe, \emph{Convex Optimization}.\hskip 1em plus 0.5em
  minus 0.4em\relax Cambridge, U.K.: Cambridge Univ. Press, 2004.

\end{thebibliography}

\end{document}